# Large remnant polarization in a wake-up free $Hf_{0.5}Zr_{0.5}O_2$ ferroelectric film through bulk and interface engineering

Alireza Kashir*, Hyung Woo Kim, Seungyeol Oh, Hyunsang Hwang*

Center for Single Atom–based Semiconductor Device and Department of Materials Science and Engineering, Pohang University of Science and Technology (POSTECH), Pohang, Republic of Korea

*kashir@postech.ac.kr, hwanghs@postech.ac.kr

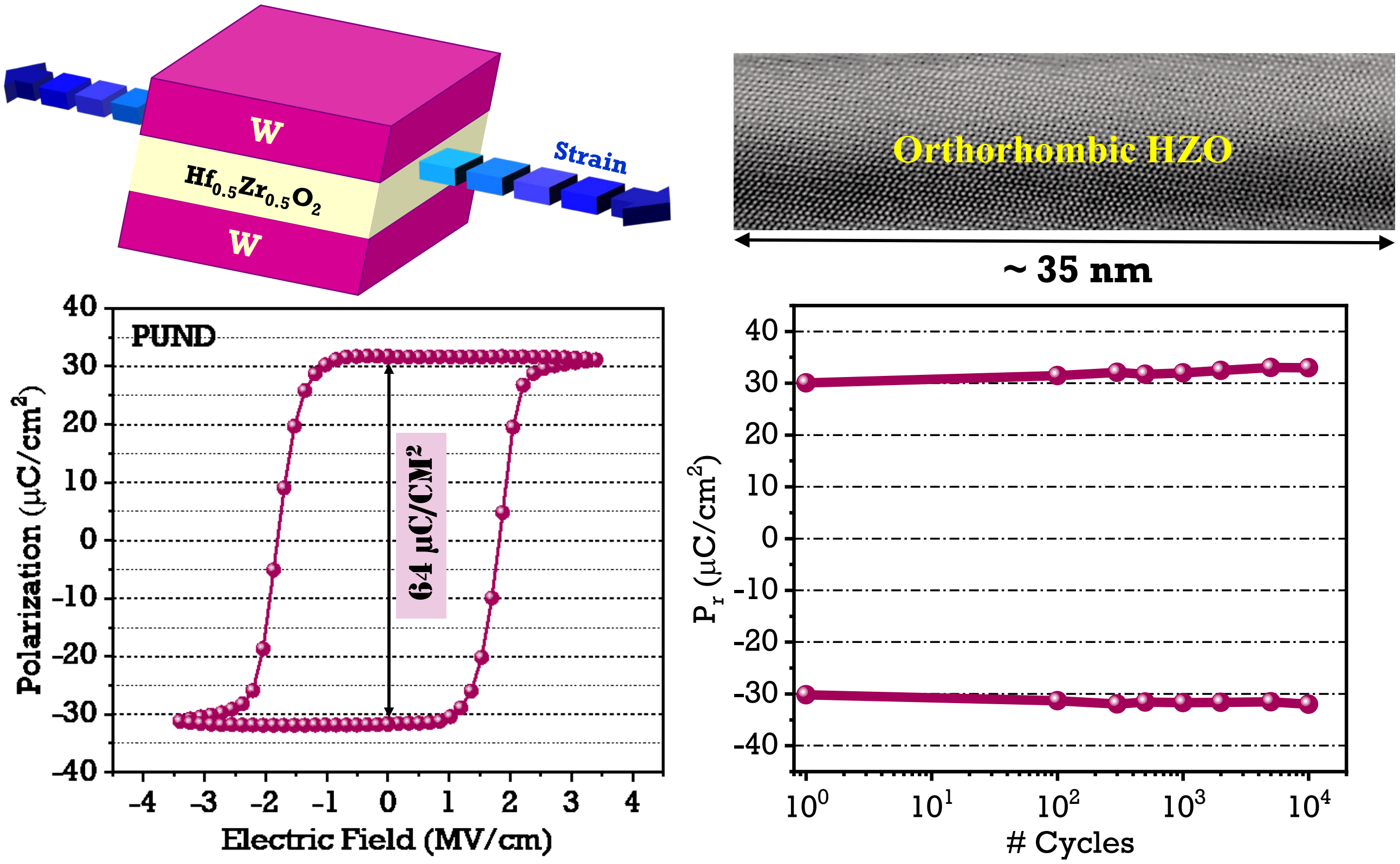

**Abstract**

A wake-up free $Hf_{0.5}Zr_{0.5}O_2$ (HZO) ferroelectric film with the highest remnant polarization ($P_r$) value to-date was achieved through tuning of the ozone pulse duration, the annealing process, and the metal/insulator interface. The ozone dosage during the atomic layer deposition of HZO films appears to be a crucial parameter in suppressing the mechanisms driving the wake-up effect. A tungsten capping electrode with a relatively low thermal expansion coefficient enables the induction of an in-plane tensile strain, which increases the formation of the orthorhombic phase while decreasing the formation of the monoclinic phase during the cooling step of the annealing process. Therefore, increasing the annealing temperature $T_A$ followed by rapid cooling to room temperature resulted in a substantial increase in the $2P_r$ value (≈ 64 μC/cm$^2$). However, the leakage current increased considerably, which can affect the performance of metal-insulator-metal (MIM) devices. To reduce the leakage current while maintaining the mechanical stress during thermal annealing, a 10 nm Pt layer was inserted between the W/HZO bottom interface. This resulted in a ~ 20-fold decrease in the leakage current while the $2P_r$ value remained almost constant (~ 60 μC/cm$^2$). The increase in barrier height at the Pt/HZO interface compared to that of the W/HZO interface coupled with the suppression of the formation of interfacial oxides ($WO_x$) by the introduction of a Pt/HZO interface serves to decrease the leakage current.

## Introduction

Since the discovery of ferroelectric properties in $HfO_2$-based materials [1], numerous technological advantages, such as a simple structure, a strong binding energy between the oxygen and transition metal ions, a wide bandgap (~ 5.3–5.7 eV), and compatibility with current complementary metal oxide semiconductor technologies, have arose in the field. This has led to extensive research focusing on their potential diverse applications, such as ferroelectric memory, ferroelectric field effect transistors (Fe-FETs), pyroelectric sensors, and energy harvesters.

Crystalline $HfO_2$ adopts a monoclinic ($P2_1/c$) structure (m-phase) under ambient conditions [2]. At ~ 1700 ºC, a martensitic phase transition results in the formation of a tetragonal ($P4_2/nmc$) phase (t-phase), and when the temperature is increased further to ~ 2200 °C, a diffusionless t-phase to cubic ($Fm\overline{3}m$) phase (c-phase) transition takes place [2]. It was shown that these transition temperatures can be significantly altered through doping, mechanical stress, or surface modification. These high-temperature phases are achievable in $HfO_2$ thin films at room temperature (RT) [3-14]. However, none of these crystal phases were found to be ferroelectric. Rather, a non-centrosymmetric polar orthorhombic ($Pca2_1$) phase (o-phase), which is extremely close in free energy to the equilibrium nonpolar phases, is believed to be the structural origin of the ferroelectric properties observed in $HfO_2$ based thin films [2, 15]. Therefore, adjusting the experimental conditions, via strain or dopants for example, may stabilize this specific polar polymorph. Increasing the remnant polarization value ($P_r$), which has a direct relation to the fraction of ferroelectric o-phase, has always been one of the main concerns in the $HfO_2$-based ferroelectric devices. During the last decade, numerous theoretical and experimental studies have been conducted to investigate the effect of influential parameters, e.g., substrates, dopants, electrodes, annealing condition, film thickness, and defects, on the stabilization of the non-centrosymmetric o-phase and to obtain a strong ferroelectric device by suppression of any parasitic (*i.e.,* c, t and m) phases. This becomes even more crucial when the device is used for the ferroelectric tunnel junctions as the tunneling electroresistance (TER) can be substantially affected by the switching polarization [16]. Recently, Park *et al.* reviewed the defect chemistry in fluorite-structure ferroelectrics and previous experimental literature reports on ferroelectric doped $HfO_2$ films were summarized [17]. Hence, we recommend reading this valuable article [17].

The transition of the t- to the o-phase, rather than the m-phase, under the confinement of a capping layer, is energetically favorable considering the large volume expansion of the crystal (~ 5%) during the later transition, where a decrease in volume of ~ 1% occurs during this t- to o-phase transition [8]. Therefore, under certain conditions (the given Substrate/Electrode/$Hf_{0.5}Zr_{0.5}O_2$ stack), metal electrodes with a relatively low thermal expansion coefficient (TEC) compared to $Hf_{0.5}Zr_{0.5}O_2$ (HZO), which induce in-plane *tensile* strain on HZO film during the cooling step of thermal annealing, appeared to be the best capping layers (or top electrode) to further facilitate the formation of o-phase by the suppression of the twining mechanism, which is responsible for m-phase formation [18]. Moreover, in-plane tensile strain stresses on the c-axis of the t-phase favor the phase transition from the t-phase to the o-phase. Because, the longest axis of the t-phase should become longer, and the shorter axes should become shorter to transform it to the o-phase [19]. Therefore, metals with a relatively low TEC not only suppress the formation of the m-phase, but also facilitate the t- to o-phase transition [18]. The local in-plane strain $\epsilon_T$ applied to the HZO film due to the TEC lattice mismatch with the capping electrode can be calculated by

$$\epsilon_{\mathrm{T}}(T_A) = \int_{RT}^{T_{\mathrm{A}}} (\alpha_{\mathrm{film}} - \alpha_{\mathrm{Cap}})\, dT \qquad (1)$$

where $\alpha_{\mathrm{film}}$ and $\alpha_{\mathrm{Cap}}$ are the TEC of the film and capping materials, respectively, and $T_{\mathrm{A}}$ is the post-annealing temperature. Therefore, if $\alpha_{\mathrm{film}}$ and $\alpha_{\mathrm{Cap}}$ are constant in the temperature range of the annealing process, the calculation is as follows:

$$\epsilon_{\mathrm{T}}(T_A) = (\alpha_{\mathrm{film}} - \alpha_{\mathrm{Cap}})\Delta T \qquad (2)$$

indicating the effect of the annealing temperature on the applied mechanical stress of the film during the cooling process to RT.

Among the most widely used metal electrodes, tungsten shows the lowest TEC (~ $4.5 \times 10^{-6}$ $K^{-1}$) and facilitates the formation of the o-phase [18]. Therefore, an appropriate annealing process under the confinement of a tungsten electrode can provide an additional driving force for the favorable t- to o-phase transition, consequently increasing the ferroelectric polarization. In contrast, it should be noted that increasing the annealing temperature causes excessive grain growth, leading to the stabilization of the m-phase and subsequent loss of ferroelectric properties [9, 14]. Therefore, as the annealing temperature increases, the annealing time should be reduced accordingly to prevent the undesirable grain growth at elevated temperatures. Another undesirable effect of increasing the annealing temperature is the formation of an oxygen-deficient dead layer at the metal/oxide interface. The result is an accumulation of oxygen vacancies due to the thermal activation of the diffusion process leading to an abrupt increase in leakage current causing serious obstacles for the operation of the device [20]. Therefore, despite the great advantages of increasing the annealing temperature during rapid thermal annealing (RTA), the negative aspects should be considered in order to fabricate a device with excellent ferroelectric properties.

In addition to tungsten, platinum is another excellent electrode material due to its chemical stability when adjacent to $HfO_2$-based thin films [21] and its high work function [22]. The former can prevent oxygen out-diffusion from the $HfO_2$ film (which causes oxygen-deficient regions in the film), and the later prevents electron injection into the film [22] due to the high Schottky barrier at the Pt/HZO interface. However, on the negative side, Pt has a high TEC of $\sim 9 \times 10^{-6}$ $K^{-1}$, which is almost the same as that of HZO [18, 23], which may be undesirable when compared with tungsten electrodes [18]. Moreover, the experimental results showed that the Pt electrode acts as a catalyst during the forming gas annealing for H-incorporation, resulting in a substantial decrease in the $2P_r$ value [24].

The primary role of ozone pulses during the atomic layer deposition (ALD) of $HfO_2$-based thin films is to infuse oxygen into the Hf and Zr layers, leading to the formation of $HfO_2$ and $ZrO_2$. Thus, the pulse length is a crucial parameter for providing enough oxygen for stoichiometric oxide formation and to avoid the generation of oxygen vacancies [25]. In addition, C–O bonds present as a result of incomplete reactions of the precursors can prevent the agglomeration of the nanoscale domains, resulting in the stabilization of the tetragonal non-ferroelectric phase [25-28]. The t-phase can subsequently transfer to the o-phase under the application of a strong electric field resulting in the wake-up effect. Therefore, the removal of the C−O bonds from the sample by applying ozone pulses with an adequate length can remarkably change the subsequent annealing behavior of the as-grown film [25, 28].

Considering these facts, in this study we adjusted the ozone dosage during the ALD growth of the HZO layers, which resulted in an almost wake-up free HZO ferroelectric film. Following this, we leveraged different configurations of tungsten and platinum electrodes under various annealing conditions to achieve a giant ferroelectric polarization ($P_r$) with a relatively low leakage current.

## Experiments

ALD was used to deposit ~ 10 nm HZO films on a 50 nm thick W bottom electrode, which had been sputtered onto a $SiO_2$/Si substrate at RT. Tetrakis (ethylmethylamido) hafnium (IV) (Hf [N-($C_2H_5$) $CH_3$]$_4$) and Tetrakis (ethylmethylamido) zirconium (IV) (Zr [N-($C_2H_5$) $CH_3$]$_4$) were used as the Hf and Zr precursors, respectively, and $O_3$ (276 g/Nm$^3$) was used as the oxidant. During the ALD process, the $HfO_2$ and $ZrO_2$ layers were deposited at a ~ 1:1 cycle ratio. Different ozone pulse lengths ranging from 2–15 s were applied for each deposition. The substrate temperature was maintained constant at 250 °C during the deposition of all films presented in this study and the precursors' temperature was kept at 90 °C. The growth rates of $HfO_2$ and $ZrO_2$ were almost identical at ~ 1 Å/cycle. After deposition, all films were capped with a 50 nm W electrode using the rf-sputtering technique at RT to fabricate W/HZO/W capacitors. The top electrode with different sizes from 30 × 30 μm$^2$ to 100 × 100 μm$^2$ was patterned using a lift-off process via the photolithography method. Finally, the W/HZO/W capacitors with different electrode areas were subjected to the annealing process under an $N_2$ atmosphere at varying temperatures ranging from 500 to 800 °C. To further investigate the interface effect, we sputtered a ~ 10 nm Pt layer between the W/HZO bottom and top electrodes to fabricate the Pt/HZO interface.

The crystal structures of the films were investigated using an X-ray diffractometer with a grazing incidence geometry (GIXRD) and high-resolution transmission electron microscopy (HRTEM). To determine the elemental composition of the films, X-ray photoelectron spectroscopy (XPS) was performed under the in depth profiling mode (by the $Ar^+$ ion sputtering) on the as-grown uncapped samples. The ferroelectric properties of the metal-insulator-metal (MIM) capacitors were measured using an LCII ferroelectric precision tester (Radiant Technologies), and the C-V and I–V characteristics were evaluated using a Keysight B1500A semiconductor device parameter analyzer. The dielectric constant $\varepsilon_r$ for each sample was extracted from the C-V measurements.

## Results and discussion

Figure 1a shows the curves of pristine polarization versus electric field (P-E) for the samples grown at different ozone pulse durations. These samples were passed through a rapid annealing process at 500 ºC for 30 sec. The samples that were grown with ozone pulses of less than 5 s exhibited a pinched hysteresis loop, while the one that was deposited with a 15 s pulse showed an open loop [28]. The evolution of the P-E loop during the electric field cycling revealed that the decrease in ozone pulse length during the deposition process results in a device with a larger wake-up effect (Figure 1b and 1c) [28]. The woken-up P-E loops presented in figures 1b and 1c demonstrate that the sample that was grown at an ozone pulse time of 5 s underwent a large wake-up effect, while that grown at ozone pulses of 15 s demonstrated ~ 90% of its woken-up $P_r$ value at the pristine state.

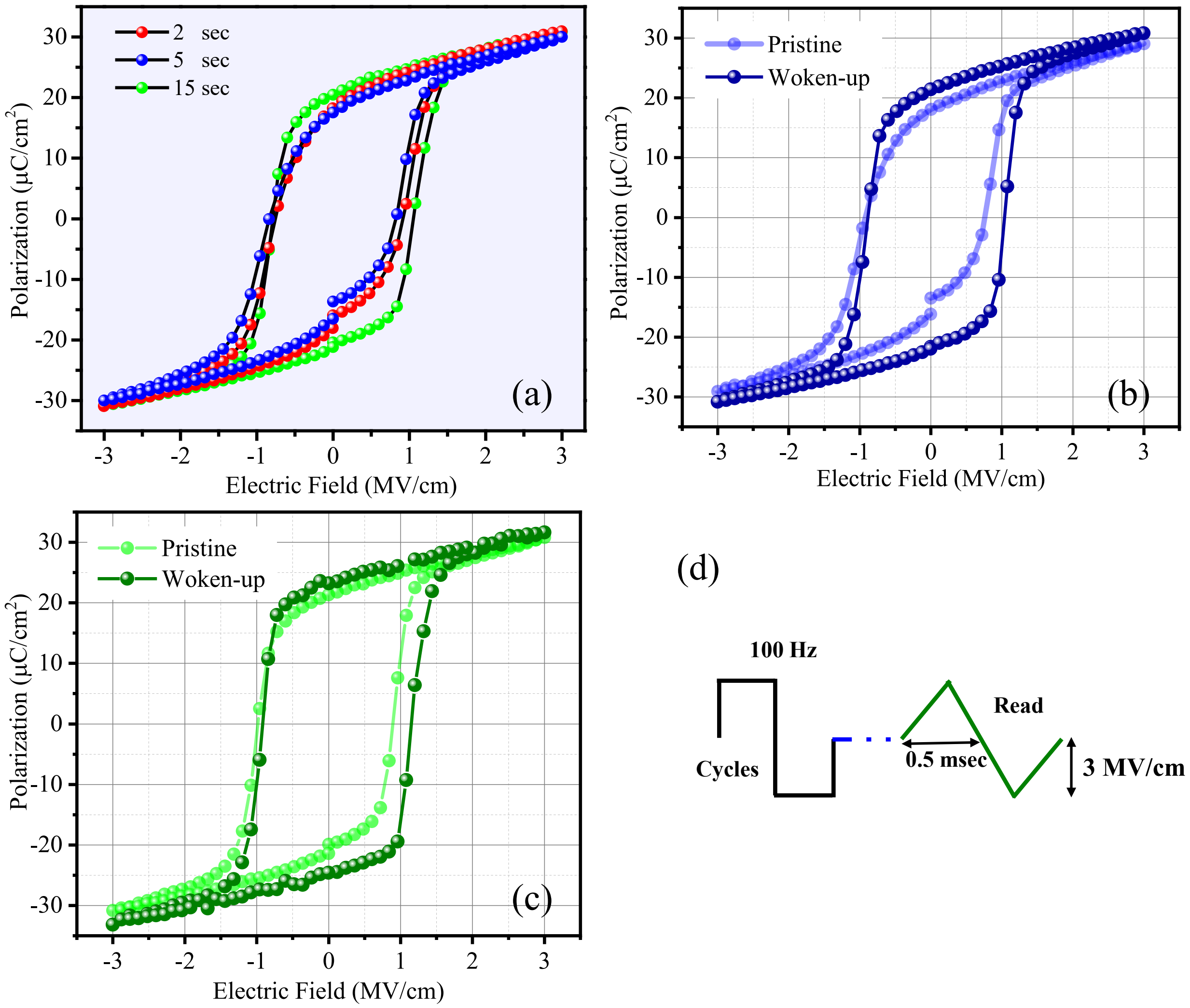

**Figure 1.** (a) The pristine P-E loops for the W/HZO/W devices fabricated at different ozone pulse times. The woken-up and pristine P-E loops for the samples grown at an ozone pulse duration of (b) 5 s and (c) 15 s. (d) The cycling and measurement sequence performed on the samples to reach woken-up P-E loops.

The woken-up $2P_r$ value for both samples were almost the same. The cycling and measurement sequence performed on the samples to reach wake-up P-E loops are presented in figure 1d. These samples underwent ~ 1000 cycles at the mentioned condition to reach their woken-up values.

This behavior was attributed to the effect of the ozone pulse on the suppression of the mechanisms that drive the wake-up effect [28]. That is, oxygen vacancies and the electric field-induced phase transition from the non-ferroelectric phases (m- and t-phase) to the ferroelectric o-phase [28-36]. A detailed study on the effect of ozone pulse length on the ferroelectric behavior of the HZO film was presented in our previous work [28]. Oxygen vacancy is believed to increase the stability of the o-phase in the $HfO_2$-based thin films [37], yet it also causes the introduction of a locally distributed inhomogeneous internal field resulting in a pinched hysteresis loop at the pristine state [29-32]. Ozone pulses at an adequate length can cause the formation of a stoichiometric HZO film, while also removing point defects during the deposition process. Moreover, ozone pulses during ALD are known to remove carbon, which may be present as a result of the incomplete reaction of the organic precursors and can substantially affect the crystallization characteristics of the samples during thermal post-annealing. Previous research has shown that incomplete carbon removal under an insufficient oxygen environment stabilizes the t-phase in annealed $HfO_2$-based thin films [25-28]. The remaining t-phase might then undergo a phase transition to the o-phase during electric field cycling, which results in wake-up behavior [35]. Therefore, an ozone pulse with an adequate length appeared to be promising for achieving a wake-up free HZO ferroelectric thin film [28].

For further investigation, XRD and XPS studies were carried out to determine the structural characteristics and to perform the elemental analysis in different samples.

The XRD results demonstrate the suppression of m-phase during the RTA process (Figure 2a). All films contained almost only t- and o-phases in this scan, possibly arising from the in-plane tensile strain applied to the HZO film during thermal annealing. In fact, no m-phase was detected in XRD scans. We should note that these observations were limited due to the instrumental resolution. Mechanical stress on the HZO film induced by the W-capping layer during the RTA is believed to provide an *additional* driving force for the t- to o-phase transition [18]. In fact, the in-plane tensile strain, which is developed during the annealing process, substantially suppresses the formation of the m-phase by preventing the necessary twin deformations required to form this phase (Figure 2b). Therefore, the TEC of the capping material substantially alters the structural features of HZO films [18]. As the ozone pulse duration increased from 2 to 15 s, a small left-shift in the XRD peak corresponding to the o-/t-phase was observed (inset of figure 2a). This shift could be due to the increase in the $\frac{o-phase}{t-phase}$ ratio resulting in an increase in the 2$P_r$ value from ~ 37 to 44 μC/$cm^2$.

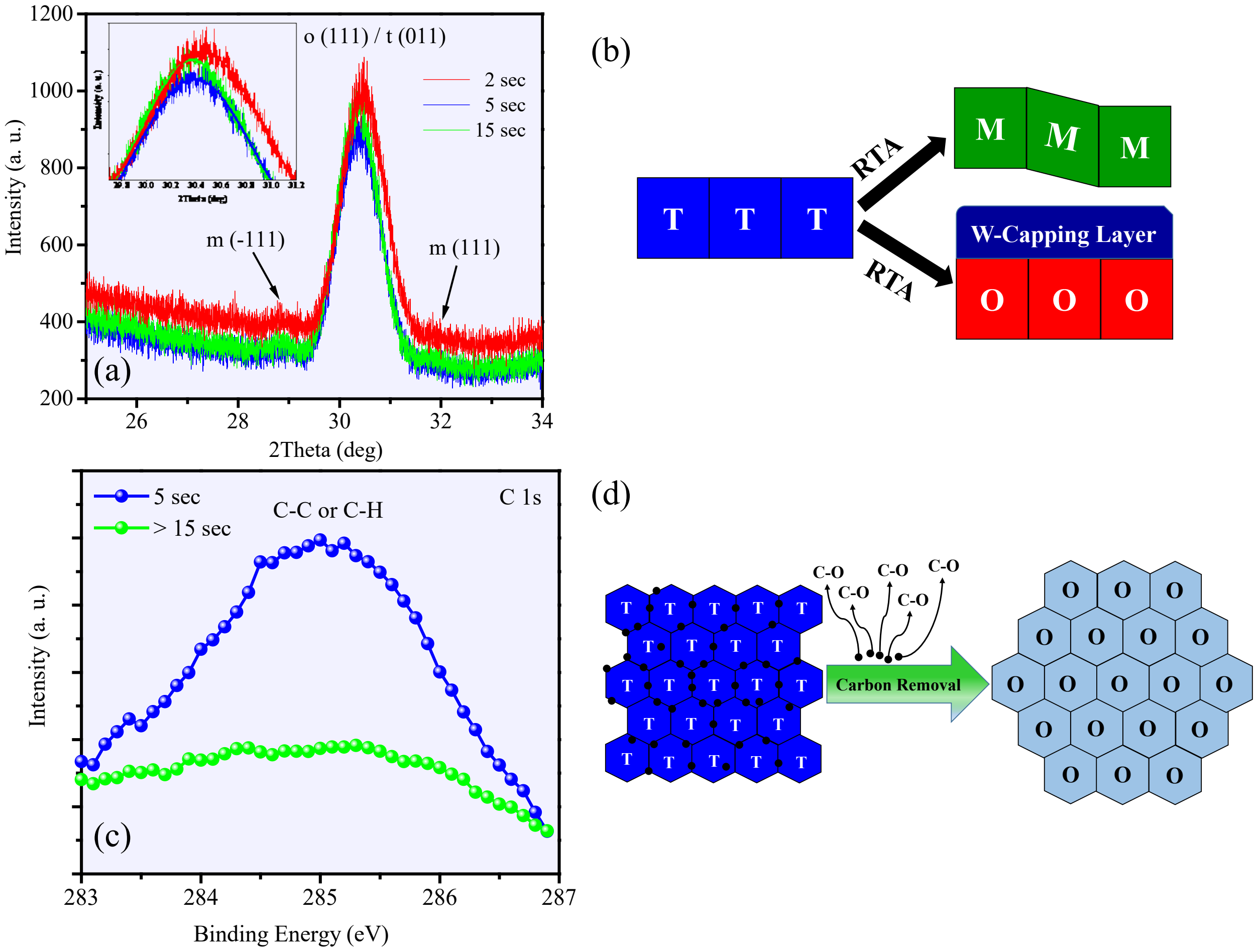

**Figure 2.** (a) XRD patterns of HZO films grown at different ozone pulse durations. (b) A schematic illustration of the phase transition from t- to m- or o-phases. The W-capping electrode suppresses the twin deformations, which are responsible for the m-phase formation. (c) C 1s XPS spectra of the as-grown HZO thin films deposited at different $O_3$ pulse times. (d) Schematic of the effect of carbon contaminations (●) on the stabilization of the t-phase in a HZO thin film. Carbon removal by a sufficient oxygen supply results in the optimum growth of grain to form the ferroelectric o-phase during the cooling step of the annealing process.

The sample deposited at the lower ozone dosage shows a higher amount of carbon (Figure 2c); the XPS measurement revealed approximately 2.1% carbon in the sample that was grown at a 5 s ozone pulse, while the samples deposited at higher $O_3$ dosages (>15 s) were almost free of carbon [28]. Carbon removal from the HZO film when deposited with a sufficient oxygen supply can facilitate grain growth during the subsequent RTA, preventing the stabilization of the t-phase during the cooling process [25-27]. A schematic (Figure 2d) shows how carbon contaminations (indicated as black circles) can pin the grain boundaries and slow the grain growth rate during the RTA process [25].

To induce a stronger in-plane mechanical stress during the cooling process of RTA according to equations 1 and 2, we increased the annealing temperature ($T_A$) to 800 °C. Notably, the annealing time at elevated temperatures is a crucial parameter leading to an excessive grain growth, which is undesirable for the formation of the o-phase. Rather, m-phase formation is preferred at elevated temperatures because of the relatively large grain size [38]. It is known that in nanoparticles of radius $r$, the surface energy ($\sigma$) produces a large internal pressure ($P=2\sigma/r$), which can facilitate the formation of the ferroelectric o-phase [9, 11, 38]. Therefore, at higher annealing temperatures, the annealing time was reduced accordingly to control the kinetics of grain growth, thereby preventing the formation of large grains, and consequently introducing a large internal pressure to the grains. Additionally, the undesirable reaction at W/HZO interface due to thermal annealing should be considered, as it substantially degrades the insulating properties of the MIM stack and causes an abrupt increase in leakage current. Continuing to probe the effect of $T_A$, annealing at 700 °C for 5 s on the W/HZO/W stacks which was fabricated under 15 s ozone dosage, resulted in a remarkable increase in the pristine $2P_r$ value to 63 μC/cm$^2$ (Figure 3a). This increase is likely due to the large mechanical stress induced by the rapid cooling from 700 °C to RT. The wake-up $2P_r$ reached ~ 65 μC/cm$^2$, which is the largest value reported for Zr-doped $HfO_2$ thin films. Previously, Schroeder *et al.* achieved a $2P_r$ value of 55 μC/cm$^2$ for the ~ 12% La-doped $HfO_2$ film which was attributed to the stronger 002 texture compared to other films [39-40], and recently Zacharaki *et al.* [41] achieved $2P_r$ ~ 60 μC/cm$^2$ for the HZO film grown on Ge substrate. Here, we adjusted the device structure using W electrodes, annealing conditions and oxygen dosage to achieve a large $2P_r$ value of ~ 65 μC/cm$^2$ in the $HfO_2$-based conventional MIM stack (W/HZO/W) deposited on the $SiO_2$/Si substrate. Annealing above 720 °C caused a large increase in leakage current, and thus the capacitors were electrically short-circuited.

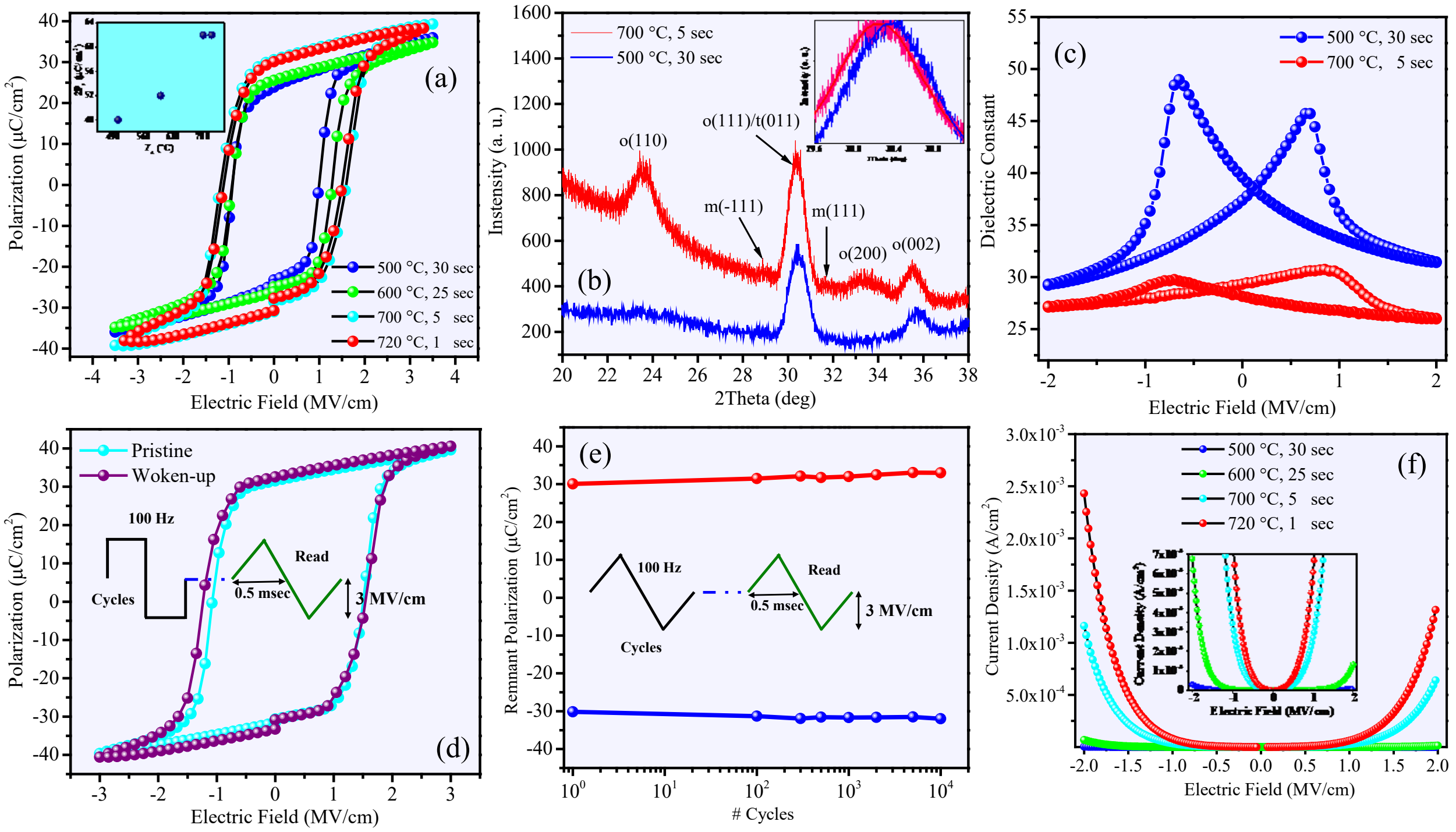

**Figure 3.** (a) Pristine P-E curves, (b) XRD patterns and (c) dielectric constant vs applied bias electric field of W/HZO/W capacitors deposited with 15 s ozone pulse durations and annealed at different temperatures. (d) The pristine and woken-up (after 1000 cycles) P-E loops and (e) the change of $P_r$+ and $P_r$- versus the number of electric field cycles for the W/HZO/W sample annealed at 700 ºC for 5 s. (f) Current density versus electric filed of the W/HZO/W capacitors deposited at 15 s ozone pulse durations and annealed at different temperatures. The inset of (a) shows the values of $2P_r$ versus annealing temperature. The inset of (f) shows the magnified areas of the relatively low leakage currents.

To investigate the origin of this significant increase in $2P_r$, the structural features of the samples annealed at 500 °C and 700 °C were compared. The XRD results presented in the figure 3b showed a noticeable left-shift of the o/t-phase characteristic peak for the sample annealed at 700 °C, indicating a further increase in the fraction of the o-phase (Inset of figure 3b). In fact, our investigation showed that the sample annealed at 700 °C is approximately free of t-phase. The o/t Bragg peak was located at ~ 30.4º, which is the characteristic peak of o (111) planes. In both samples, the formation of m-phase was suppressed and there is no observable m-phase peak in XRD patterns. A simultaneous substantial increase (~ 150%) in the coercive field ($E_c$) was observed. The appearance of o (110) peak and more intensified o (002) peak after annealing at 700 °C compared to the other sample might be the cause of substantial increase in the remnant polarization. It was demonstrated that the better orientation of the polar 002 axes toward the electric filed is the origin of the outstanding polarization value in La-doped $HfO_2$ [40]. In this work, by adjusting MIM configuration (using W top and bottom electrodes), ozone dosage and annealing condition, the device shows an XRD pattern almost similar to the La-doped $HfO_2$ presented in reference [40]. Appearance of the O (110) crystals might be due to the better crystallization at the elevated temperatures which causes less amount of amorphous HZO inside the deposited film. The dielectric constant calculated from C-E measurements was decreased as the annealing temperature increased (Figure 3c). This is in agreement with the reduction of t-phase and the increase of o-phase. The t-phase in HZO has larger dielectric constant compared to the o-phase [36]. The $E_c$ of the $HfO_2$-based ferroelectric thin films is believed to remain constant, becoming independent of the film thickness as it decreases below 20 nm [42, 43]. A possible reason for the increase of the coercive field as the annealing temperature increases might be the oxidation of W at W/HZO interface. Consequently, a large number of oxygen vacancies in both HZO and $WO_x$ might be generated. The increase of the leakage current through the W/HZO/W samples annealed at higher temperatures reflects this fact (Figure 3f). The defective interface can decelerate the nucleation and growth rate of the reversal domains as was proven to be the main mechanism of the polarization switching phenomenon in HZO thin film [44]. This, in turn, leads to an increase in the coercive field. Domain wall needs to nucleate a bulge that is free of trapping defects to break free from the attractive potential well (Figure S2).

The woken-up P-E loop obtained after 1000 cycles at 3 MV/cm from the sample annealed at 700 °C for 5 s showed a small change in the $2P_r$ value (Figure 3d), which indicates that the W/HZO/W ferroelectric device is almost wake-up free. The suppression of m-phase and the reduction of t-phase revealed through XRD results (Figure 3b), due to a proper annealing condition might decrease the wake-up effect in $HfO_2$-based thin film [45]. The endurance measurement using triangle waveform electric filed with the frequency of 100 Hz clearly indicates an almost wake-up free device until its break-down cycle (Figure 3e). However, the I-V measurement revealed a significant increase in the leakage current of the samples annealed at higher temperatures (Figure 3f), which is believed to have an undesirable effect on the stability of MIM devices. Even decreasing the annealing time from 30 to 1 s did not prevent the large increase in leakage current, revealing the critical role of the annealing temperature in the activation of the conduction mechanisms in W/HZO/W capacitors. The origin of the leakage current may be a result of the formation of tungsten oxides ($WO_x$) at the W/HZO interface, whereby the thermal energy provided by the deposition and annealing processes can facilitate the formation of $WO_x$, subsequently leading to an oxygen-deficient interface. The formation of grains during the thermal process can be another possible mechanism for the increase in leakage current.

To improve the current behavior of the HZO capacitor, we added an ~ 10 nm Pt layer between the W and HZO top and bottom interfaces in order to increase the potential barrier at the metal/insulator interface and decrease the possible formation of an interfacial layer. Among the most widely used metal electrodes, Pt has the highest work function (~ 5.65 eV) [22], which results in a high potential barrier at the interface with the insulating film (Figure 4a). Fillot *et al*. [21] studied the stability of Pt electrodes in connection with $HfO_2$ films and stated that the Gibbs free energy shown as follows: $\Delta G$ of $2Pt + HfO_2 \rightarrow Hf + 2PtO$ was >1000 kJ/mol at 500 °C. This emphasizes the stability of Pt metal when in contact with the $HfO_2$ film. Therefore, the possibility of oxygen out-diffusion from the film that suppresses the formation of an interfacial layer during the thermal deposition and annealing processes is reasonably low when compared to the W/HZO interface. Since the formation energy of $ZrO_2$ is almost the same as that of $HfO_2$ [46], we assumed that this condition also corresponds to the case of a (Hf, Zr) $O_2$ solid solution. Moreover, the high work function of Pt, compared to that of W [22], makes it a suitable electrode to prevent electron injection into the dielectric film through the electrode/film interface via a Schottky mechanism. Therefore, the two important mechanisms that yield a substantial increase in leakage current can be suppressed by the introduction of a thin Pt layer at the W/HZO interface. Figure 4a compares the relative Schottky barriers between the Pt/HZO and W/HZO interfaces. The I-V measurements of different capacitors were performed at 200 °C, at which the thermal Schottky emission is likely to occur [47]. The intercepts of the Schottky plots show different barrier heights for W/HZO/W and Pt/HZO/Pt capacitors, being higher for the Pt/HZO interface (Figure 4b). Therefore, inserting a thin Pt layer between the W/HZO interface may substantially reduce the leakage current through the MIM stacks.

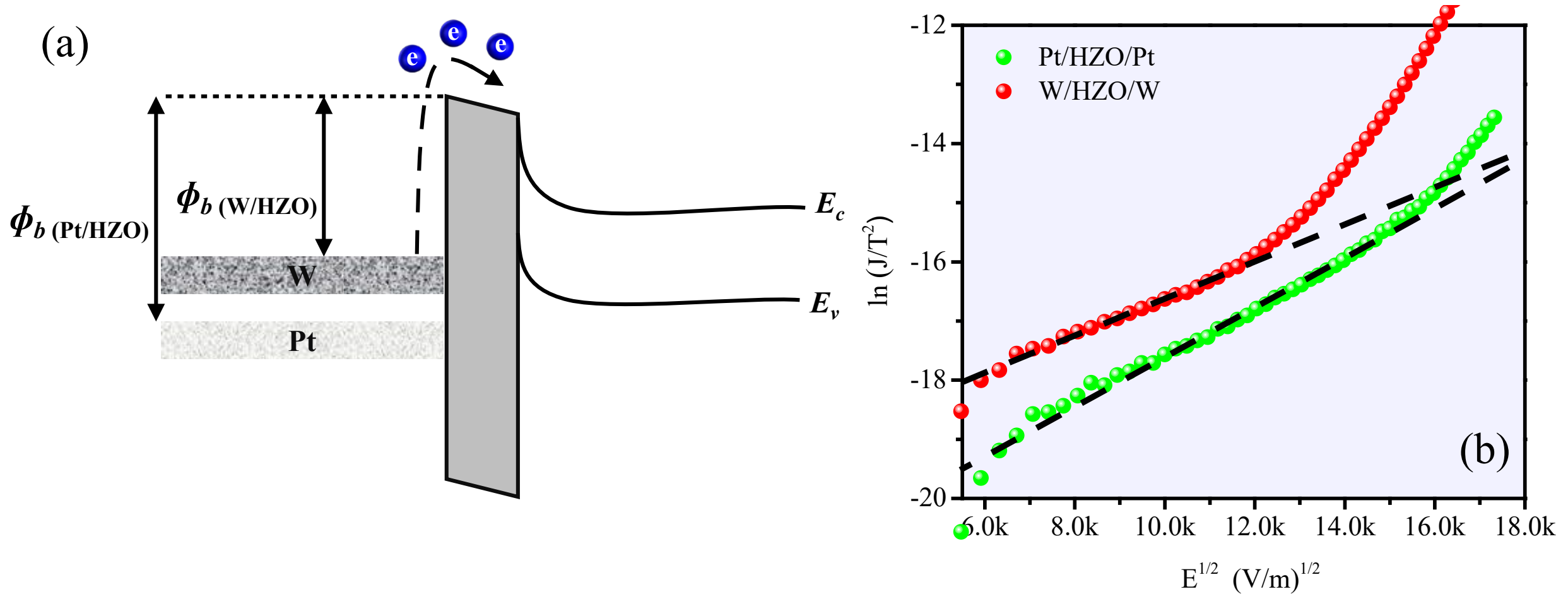

**Figure 4.** (a) Schematic illustration of the Schottky plot at the interface of W/HZO and Pt/HZO, (b) The linear Schottky relationship between ln ($J/T^2$) and $E^{1/2}$ for different capacitors.

The I-V measurement showed a substantial decrease in the leakage current of the MIM capacitor as a result of the introduction of a Pt layer between the W electrodes and the HZO films (Figure 5a). The leakage current at 2 MV/cm decreased by ~ 2 orders of magnitude through the introduction of a Pt layer, but using Pt as a capping electrode revealed a considerable degradation in the ferroelectric polarization (Figure S1) [18]. The degradation of ferroelectric polarization was attributed to the TEC of Pt ~ $9\times10^{-6}$ $K^{-1}$, which is almost same as that of HZO [18, 23]. According to equation 1, a negligible in-plane strain should be applied to the HZO film during the cooling process, which is not suitable for driving the transition from the t- to o-phase. Therefore, adding a 10 nm Pt layer at the HZO/W top interface has a strong impact on the mechanical stress applied to the HZO layer during the RTA. To optimize the advantages of both the W and Pt electrodes in order to improve the ferroelectric properties of the HZO thin film, we fabricated a W/Pt/HZO/W capacitor that can maintain the mechanical stress condition at the top interface and control the bottom interface properties, which is thermally exposed to oxygen during ALD. The I–V measurement showed an almost 20-fold decrease in leakage current when compared to that of W/HZO/W.

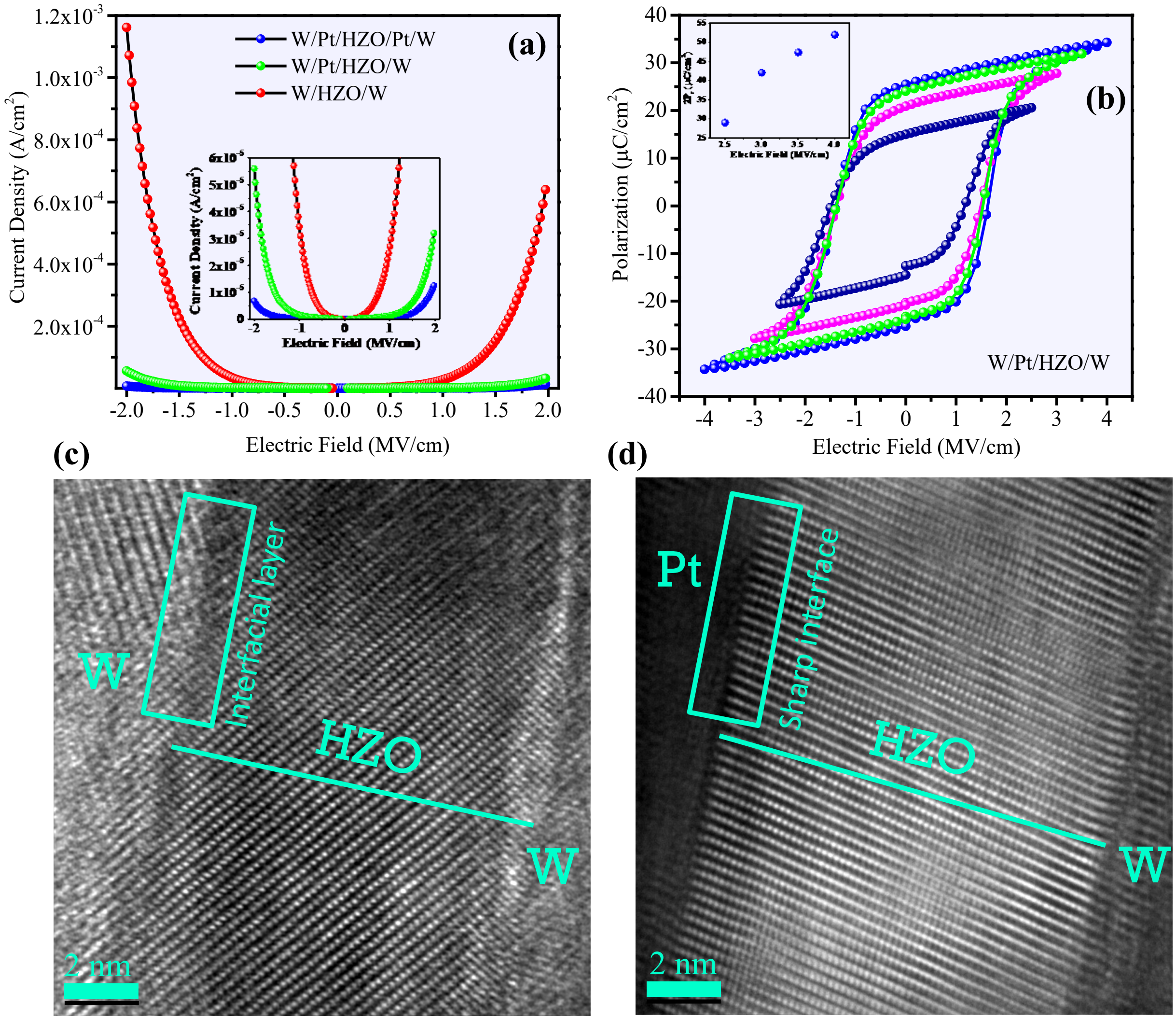

**Figure 5.** (a) The current density versus electric field of different capacitors fabricated on an $SiO_2$/Si substrate annealed at 700 ºC for 5 sec. The inset shows the magnified areas of relatively low leakage currents. (b) The pristine P-E loops of W/Pt/HZO/W device. Inset shows the $2P_r$ value versus applied electric filed. Interface study using high-resolution transmission electron microscopy (HRTEM). Cross-sectional HRTEM images of (c) W/HZO/W, (d) W/Pt/HZO/W devices.

We should note that the W/Pt/HZO/W device reaches its saturated P-E loop at ~ 4 MV/cm (Figure 5b) while the W/HZO/W device showed the saturated P-E loop at ~ 3.5 MV/cm (Figure S3). This might be due to the oxidation of W electrode, which can pull out oxygen from HZO layer and reduce the effective thickness of insulating HZO film. The cross-sectional HRTEM images of these two samples are compared in the figures 5c and 5d. In fact, at W/HZO interface some interfacial regions were formed which might be due to the oxidation of W electrode during the device fabrication process. On contrary, the Pt/HZO demonstrates a sharp interface. The formation of interfacial deficient layer may reduce the thickness of insulator in an MIM capacitor. Therefore, under a given applied voltage the calculated electric filed ( $E = \frac{V_{app}}{t_{insulator}}$ ) would be underestimated. In this work, the actual electric field applying to the HZO layer in W/HZO/W device might be higher than that of W/Pt/HZO/W under a same applied voltage. The reported electric fields are the *nominal values* based on the assumption that the thickness of HZO is 10 nm in both devices.

The pristine P-E loop of W/Pt/HZO/W shows a $2P_r \sim 52$ μC/cm$^2$ which undergoes a wake-up effect of 7 μC/cm$^2$ to reach the woken-up $2P_r = 59$ μC/cm$^2$. The endurance measurement using a triangle waveform electric filed with an amplitude of 3 MV/cm and frequency of 1 kHz revealed an order of magnitude increase by adding a thin Pt layer between W and HZO bottom interface (Figure 6a). To correct for leakage and possible contributions from parasitic charges, the positive up-negative down (PUND) method was used. This consists of applying a negative set pulse followed by two positive pulses (P and U) and then two negative pulses (N and D). The P-E loops obtained from the PUND method (Figure 6b) showed a $2P_r$ value of approximately 64 μC/cm$^2$ and 59 μC/cm$^2$ for W/HZO/W and W/Pt/HZO/W stacks, respectively: the highest $2P_r$ values reported for Zr-doped $HfO_2$ in the literature hitherto. The XRD result for the W/Pt/HZO/W device presented in figure 6c, shows almost the same as W/HZO/W. In case of W/HZO/W device, the o (110) peak appeared which is absent in W/Pt/HZO/W XRD pattern. This might be the origin of the different pristine $P_r$ values for these two devices. Reduction of the thickness of Pt bottom buffer layer to few nanometers might increase the direct mechanical effect of 50 nm W bottom electrode on HZO layer. Consequently, besides the reduction of oxygen deficiencies which improves the endurance of the device, the ferroelectric polarization of the W/Pt/HZO/W device remains almost the same as W/HZO/W.

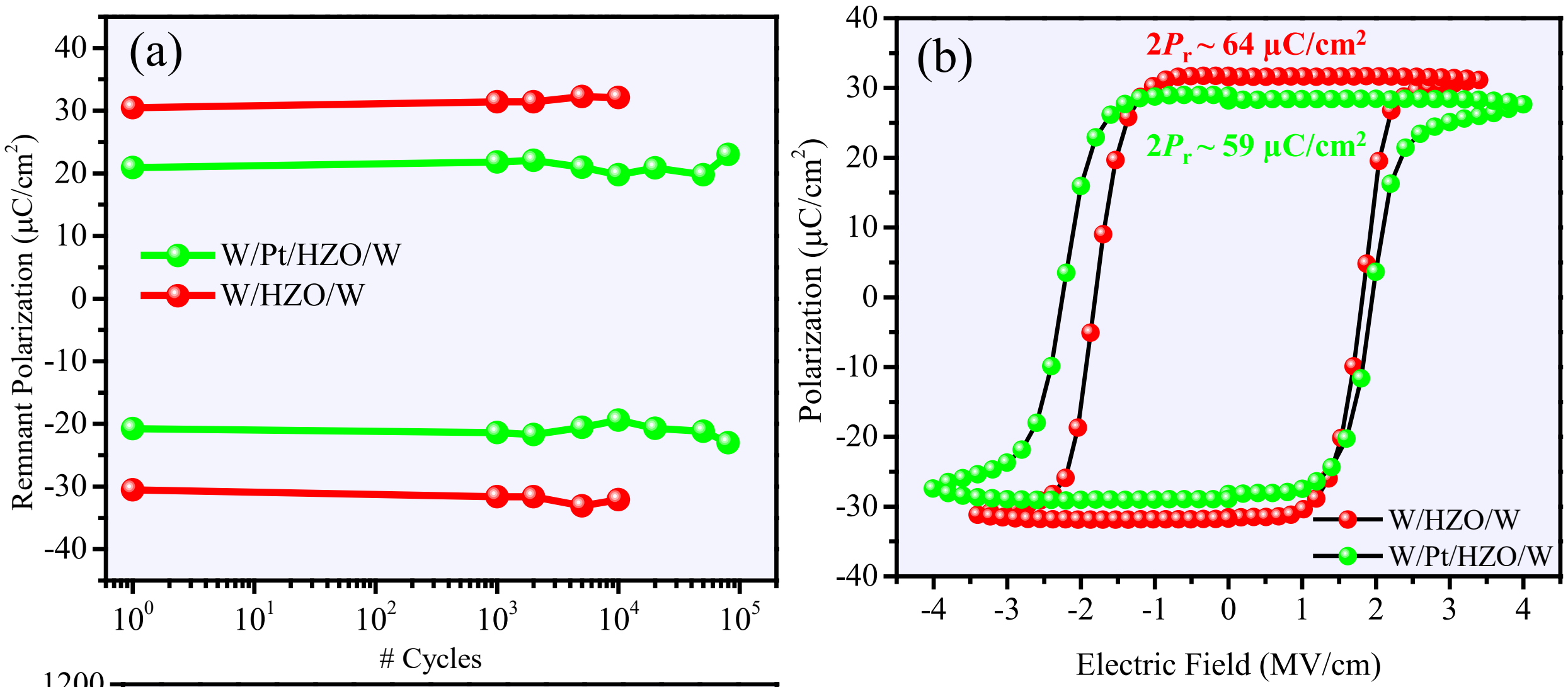

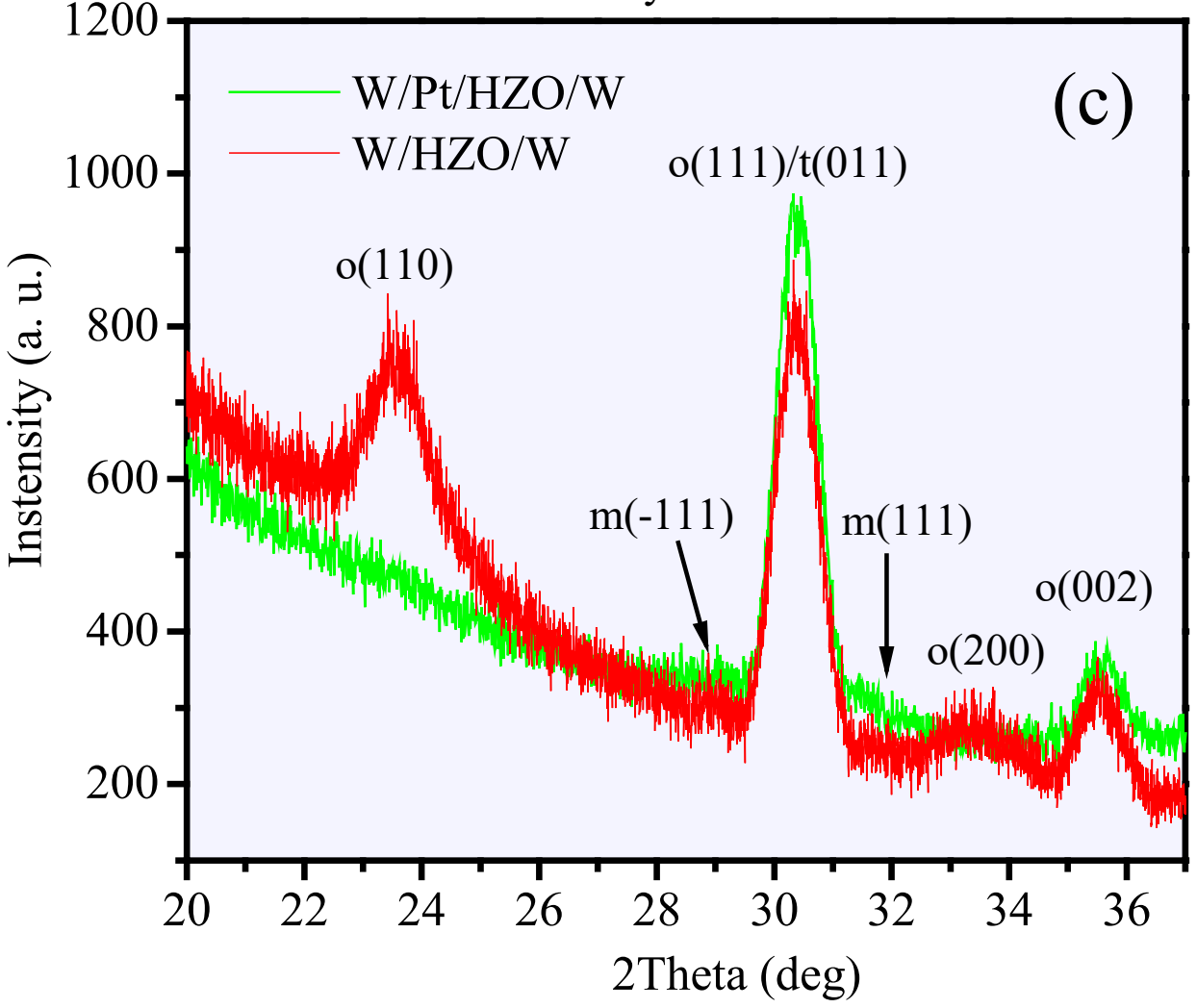

**Figure 6.** (a) The endurance measurement, (b) the corrected P-E loops obtained using the PUND method, and (c) the XRD patterns of different capacitors fabricated on $SiO_2/Si$ substrates and annealed at 700 ºC for 5 sec.

Finally, an extended scale cross-sectional HRTEM image of a W/Pt/HZO/W device is presented in figure 7. Supported with GIXRD scan of the o/t characteristic peak at $2\theta = 30.4^o$ (Figure 6c), this image reveals the formation of large orthorhombic grains [15] which could be the origin of the high $P_r$ observed in this work. Since the formation of large-grain o-phase is not energetically favorable and $HfO_2$-based thin films are supposed to contain large-grain m-phase, we can conclude that the in-plane tensile strain applied from W capping electrode during RTA has the major contribution to achieving high polarization value in this device.

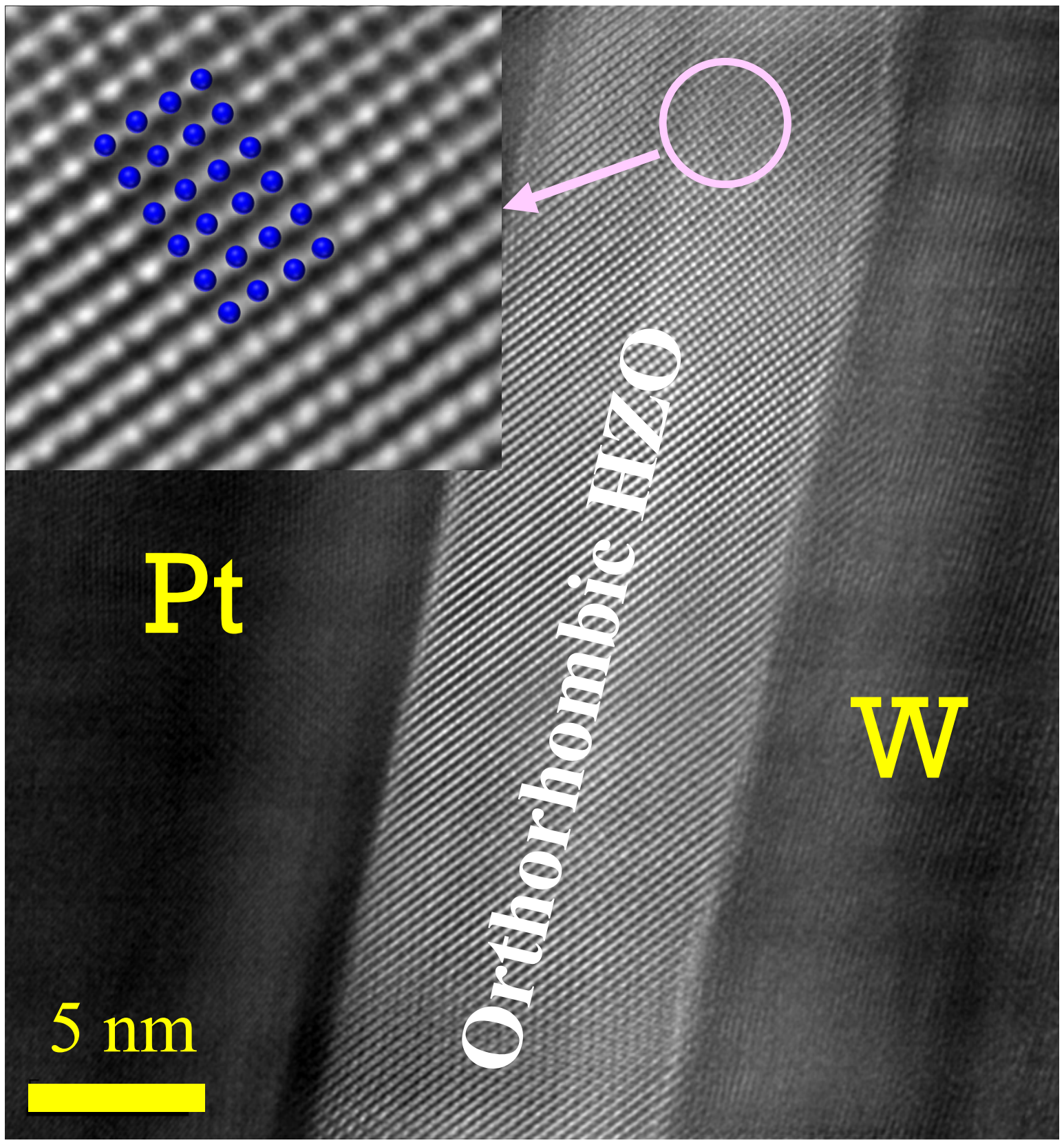

**Figure 7.** A cross-sectional HRTEM image of W/Pt/HZO/W device fabricated at 15 sec ozone pulse time and annealed at 700 ºC for 5 sec. Inset shows a magnified view of the atomic arrangement along [100] direction.

## Conclusions

By adjusting the ozone pulse length during the ALD growth of HZO and increasing the annealing temperature during RTA, a wake-up free W/HZO/W ferroelectric device with $2P_r$ value ~ 64 $\mu C/cm^2$ was obtained. This is one of the highest $2P_r$ values achieved in $HfO_2$-based ferroelectric to-date. The main contributing factor in achieving this high $2P_r$ value is the substantial suppression of the m- and t-phase in the HZO film after the RTA process, which was due to the carbon removal from the as-deposited HZO film by increase of ozone dosage and the in-plane tensile strain applied to the HZO film by the W bottom and capping electrode. In fact, the GIXRD data revealed no parasitic phase in the HZO films and HRTEM showed large orthorhombic grains, which could be the cause of high $2P_r$ value observed in this work. Moreover, the appearance of o (110) and more intensified o (002) Bragg peak which was believed to be the main reason of the large $2P_r$ value of La-doped $HfO_2$ film, might cause a huge increase in $2P_r$ value of HZO film annealed at 700 ºC. The insertion of a thin Pt layer between the W/HZO bottom interface reduced the leakage current by increasing the Schottky barrier and preventing the possible oxidation of the W electrode adjacent to the HZO film which was confirmed by HRTEM study. The top W/HZO interface was retained to provide a strong in-plane tensile stress, which is desirable to induce the formation of the ferroelectric o-phase in the HZO film. Therefore, strong ferroelectric properties were achieved by adjusting the bulk and interfacial conditions of the W/HZO/W capacitors.

**Acknowledgements**

This work was supported by the National Research Foundation of Korea funded by the Korea government (MSIT), Grant No. NRF-2018R1A3B1052693. The authors thank Ms. Yeganeh Manouchehr, Dr. Writam Banerjee and Dr. Nikam for helpful discussions.

**Supporting Information**

P-E curves of W/Pt/HZO/Pt/W and W/HZO/W, A schematic illustration of the nucleation stage of the reversal domains in W/HZO/W device, P-E curves of W/HZO/W and W/Pt/HZO/W devices annealed at 700 ºC for 5 sec